\documentclass[prd,twocolumn,nofootinbib,superscriptaddress,amsmath,amssymb]{revtex4-1}
\usepackage[utf8]{inputenc}
\usepackage{mathtools}
\usepackage{graphics}
\usepackage{graphicx}
\usepackage{dcolumn}
\usepackage{bm}
\usepackage{dsfont} 
\usepackage{amsmath,amssymb}
\usepackage{hyperref}
\usepackage{tabularx}
\usepackage{epstopdf}
\usepackage{tensor}
\usepackage{mathrsfs}
\usepackage[normalem]{ulem}
\usepackage[usenames]{color}
\hypersetup{
    colorlinks=true,
    linkcolor=Fuchsia,
    filecolor=magenta,      
    urlcolor=Fuchsia,
    citecolor=Fuchsia
}
\usepackage[dvipsnames]{xcolor}
\usepackage{mathrsfs}
\usepackage{comment}

\newcommand{\beq}{\begin{equation}}
\newcommand{\eeq}{\end{equation}}
\newcommand{\be}{\begin{equation}}
\newcommand{\ee}{\end{equation}}

\newcommand{\Brown}{Brown Theoretical Physics Center and Department of Physics, Brown University, 182 Hope Street, Providence, Rhode Island, 02903}

\begin{document}

\title{Higher Spin Dark Matter}

 \author{Stephon Alexander}
\affiliation{\Brown}
\affiliation{Center for Computational Astrophysics, Flatiron Institute, New York, NY 10003, USA}

 \author{Leah Jenks}
 \affiliation{\Brown}
 
  \author{Evan McDonough}
\affiliation{Kavli Institute for Cosmological Physics and Enrico Fermi Institute, The University of Chicago, Chicago, IL 60637, USA}

\begin{abstract} 
Little is known about dark matter beyond the fact that it does not interact with the standard model or itself, or else does so incredibly weakly. A natural candidate, given the history of no-go theorems against their interactions, are higher spin fields. Here we develop the scenario of higher spin (spin $s>2$) dark matter. We show that the gravitational production of superheavy bosonic higher spin fields during inflation can provide all the dark matter we observe today. We consider the observable signatures, and find a potential characteristic signature of bosonic higher spin dark matter in directional direct detection;  we find that there are distinct spin-dependent contributions to the double differential recoil rate, which complement the oscillatory imprint of higher spin fields in the cosmic microwave background. We consider the extension to higher spin fermions and supersymmetric higher spins.
\end{abstract}

\maketitle

\section{Introduction}

The precise identity of dark matter (DM) remains a mystery, despite decades of theorizing and detection efforts. Observations suggest that the dark matter does not interact with the standard model, or does so extremely weakly, creating significant room for model builders. Taken at face value, the observational evidence is at odds with the conventional origin story of dark matter, namely a thermal history, wherein the dark matter was initially in a state of thermal equilibrium with the standard model, sustained by interactions.

There are now many alternative dark matter origin stories. A particularly compelling possibility, by virtue of its simplicity, is the genesis of dark matter via gravitational particle production (GPP) in the early universe \cite{Kolb:1998ki,Chung:2001cb,Chung:1998bt,Chung:1998rq,Kolb:2007vd}, e.g., during cosmic inflation.  A generic feature of inflation is that the exponential expansion acts as a gravitational amplifier for particle production.  While many of these particles would be redshifted, some, depending on their intrinsic properties such as mass and spin, can survive as a relic after inflation ends.  This idea was introduced in the context of superheavy `WIMPzilla' dark matter, characterized by dark matter masses greater than the Hubble scale at the end of inflation ($m_{\rm DM} > H$) \cite{Kolb:1998ki, Chung:1998zb,Chung:2001cb,Chung:1998bt,Chung:1998rq,Kolb:2007vd}, and has since been explored in a variety of cosmological contexts (see e.g. \cite{Benakli:1998ut, Kannike:2016jfs, Kolb:2017jvz, Li:2020xwr, Alexander:2020tsf, Kolb:2020fwh}).

On the other hand, inflation is known to exhibit a UV sensitivity \cite{Copeland:1994vg}, motivating the search for a UV completion of inflation in theories beyond the standard model, such as string theory.  In this context we may ask which states could be generically produced in an inflationary model that is connected to string theory. That massive higher spin particles are a natural consequence of string theory, and that generic cosmological inflation models induce particle production, is suggestive of the potential implications for connections to other physics, namely the dark matter problem. Moreover, the Higuchi bound on the mass of higher spin fields, $m^2 \geq s(s-1)H^2$, which must be satisfied at all times during an early universe genesis mechanism, naturally suggests dark matter in the superheavy regime, and hence, in light of \cite{Kolb:1998ki,Chung:2001cb,Chung:1998bt,Chung:1998rq,Kolb:2007vd}, a gravitational origin of higher spin dark matter.

Despite the natural candidacy of higher spin fields as dark matter (and potentially an infinite tower of such fields), there has been little work done on investigating the feasibility of any such model beyond spin-3/2 \cite{Ding:2012sm, Chang:2017dvm, Garcia:2020hyo}, spin-2 \cite{Aoki:2017cnz, Marzola:2017lbt, Armaleo:2019gil, Armaleo:2020yml} and spin-3 \cite{Asorey:2010zz, Asorey:2015hrz},  aside from the suggestion in \cite{Kulaxizi:2014yxa}, and no work in a superheavy, gravitational production context. Additionally, there has been significant recent interest in the `cosmological collider physics' program \cite{Arkani-Hamed:2015bza, Chen:2009we, Chen:2015lza} (see also the related `cosmological bootstrap' \cite{Arkani-Hamed:2018kmz,Baumann:2019oyu,Baumann:2020dch}) of studying the imprint in the cosmic microwave background (CMB) of fields with masses heavier than the Hubble scale during inflation, see e.g. ~\cite{Hook:2019zxa,Hook:2019vcn,Kumar:2019ebj,Liu:2019fag,Wang:2019gbi,Wang:2020uic,Bodas:2020yho}. This formalism has been applied to higher spin bosons \cite{Arkani-Hamed:2015bza, Lee:2016vti, Baumann:2017jvh}, as well as higher spin fermions \cite{Alexander:2019vtb}, and supersymmetric higher spin theory (\cite{Curtright:1979uz,Gates:1983nr,Kuzenko:1993jq,Kuzenko:1993jp,Gates:1996my}) \cite{Alexander:2019vtb}. However, thus far, no connections have been made between the massive higher spin particles produced by the cosmological collider and dark matter.

In this work we consider inflationary production and cosmological implications of higher spin particles and find that they can naturally serve as 100 percent of the dark matter: Higher Spin Dark Matter (HSDM).  We consider the implications of a small interaction of HSDM with the standard model, and find a characteristic angular dependence of nuclear recoil events for direct detection experiments. This mirrors the angular dependence in the cosmic microwave background non-Gaussianity that is predicted due to the the production of higher spins during inflation \cite{Arkani-Hamed:2015bza, Lee:2016vti,Alexander:2019vtb}.

The structure of the paper is as follows: in Section \ref{II} we introduce the relevant higher spin formalism. In Section \ref{III} we calculate the gravitational production of higher spin dark matter (HSDM) and show that there is a parameter space such that higher spin particles can account for all the dark matter. In Section \ref{IV} we discuss the possibilities for directional direct detection and show that there is a spin dependent contribution to the double differential recoil rate. Lastly, in Section \ref{V} we speculate on other possible observable avenues and conclude with a discussion in Section \ref{VI}.

\section{Higher Spin Field Theory}\label{II} 

The Standard Model of particle physics comprises particles with $s=0, 1/2$ and $1$, while gravity has spin $s=2$. No fundamental particle with $s>2$ has ever been observed in nature. However, there is a long history of the study of higher spins (HS).  Beginning shortly after the advent of relativistic quantum field theory \cite{s3-1}, the theory of higher spins has been developing for a century, notably \cite{Fradkin:1987ks, Vasiliev:1990en, Vasiliev:2003ev, Curtright:1979uz, Gates:1983nr, Fradkin:1987ah, Didenko:2014dwa, Sorokin:2004ie, Rahman:2015pzl, Kessel:2017mxa}; for recent work see e.g.~\cite{Gates:2010td,Gates:2011qa,Gates:2011qb,Gates:2013ska, Gates:2013rka, Gates:2013tka, Koutrolikos:2015lqa, Gates:2017hmb,Buchbinder:2017nuc,Buchbinder:2018wwg,Buchbinder:2018gle,Buchbinder:2018nkp, Buchbinder:2018wzq,Gates:2019cnl,Buchbinder:2019esz,Buchbinder:2020yip,Buchbinder:2020rex, Afkhami-Jeddi:2018apj, Kaplan:2020ldi, Kaplan:2020tdz}. 

There are well known `no-go' theorems that significantly limit the interactions of HS particles in a self-consistent quantum field theory. Generally, such theorems make it the case that in flat space, massless HS particles are forbidden from interacting with electromagnetism or gravity\footnote{Note that both during inflation and in the present day, the universe is de Sitter space} \cite{Weinberg1964PhotonsAG, PhysRev.159.1251, Grisaru:1977kk, Aragone:1979hx, Weinberg:1980kq, Porrati:2008rm}. Two notable `no-go' theorems are Weinberg's theorem \cite{Weinberg1964PhotonsAG}, which necessitates that, in flat space, there are no long range interactions with spin greater than two, and the Coleman-Mandula theorem \cite{PhysRev.159.1251}, which demonstrates that, assuming an S-matrix and finite degrees of freedom, there can be no conserved higher spin charges associated with particles of $s>2$.

A caveat to these arguments is {\it massive} higher spin theories. The mass term explicitly breaks the higher spin gauge invariance, such that there is no conserved current, and hence no conflict with the Coleman-Mandula theorem. As such, massive higher spins are not plagued by the same restrictions due to no-go theorems \cite{Bellazzini:2019bzh}. Indeed, massive higher spin excitations are intrinsic to string theory, and comprise the Regge trajectories.  Higher spin fields have been considered in studies of inflation in string theory \cite{Noumi:2019ohm, Lust:2019lmq,Scalisi:2019gfv}, and in the AdS/CFT correspondence \cite{Fradkin:1987ks, Vasiliev:2003ev, Giombi:2010vg}. It is thought that the tensionless limit of string theory is a higher spin field theory \cite{Sagnotti:2003qa,Gaberdiel:2015mra,Ferreira:2017pgt,Eberhardt:2018plx,Gaberdiel:2018rqv}, and it has been suggested that string theory itself is a symmetry broken phase of a HS field theory \cite{Gross:1987kza, Amati:1987wq, Fradkin:1987ks, Sagnotti:2013bha}.

While the full theory of higher spins is not known, progress can be made by enumerating the irreducible representations of the spacetime symmetry group, thereby identifying the building the blocks of the theory. Although the representation theory of HS fields in a general Freidmann-Robertson-Walker spacetime is not known, the representations are known for flat space and (A)dS. 

For cosmological purposes, in particular during inflation, we may make use of the results for de Sitter space. In this context, a lower bound on the higher spin mass is given by the Higuchi bound: $m^2 \geq s(s-1) H^2$ \cite{Higuchi:1986py,Deser:2001us}. Beyond this, fields can be organized into three categories of unitary, irreducible representations of the spacetime isometry group \cite{10.2307/1968990, 10.2307/1969376}. They are the principal series: 
\beq 
\label{eq:princseries}
\frac{m^2}{H^2} \geq \left(s - \frac{1}{2}\right)^2, 
\eeq
the complementary series: 
\beq 
\label{eq:complseries}
s(s-1) < \frac{m^2}{H^2} < \left(s - \frac{1}{2}\right)^2, 
\eeq
and the discrete series:
\beq 
\frac{m^2}{H^2} = s(s-1) - t(t+1). 
\eeq
In addition,  In this work, we focus on the complementary series and principal series representations. 

The evolution of HS fields in de Sitter space was derived in \cite{Lee:2016vti}, which we summarize below. The spin-$s$ generalization of Klein-Gordon equation is the Casimir eigenvalue equation of the de Sitter group \cite{Lee:2016vti},
\begin{equation}
    \left(\Box - m^2 +  (s^2 - 2 s -2 )H^2\right)\sigma_{\mu_1 ... \mu_s}=0
\end{equation}
This is supplemented by constraint equations corresponding to transverse and traceless conditions on $\sigma$. To solve this equation, we expand the field $\sigma_{\mu_1\cdots\mu_s}$ into its different helicity components,
\begin{align}
\sigma_{\mu_1\cdots\mu_s} &= \sum_{\lambda=-s}^s \sigma^{(\lambda)}_{\mu_1\cdots\mu_s}\, .
\end{align}
A mode of helicity $\lambda$ and $n$ polarization directions can be written as,
\begin{align}
\sigma^{(\lambda)}_{i_1\cdots i_n\eta\cdots \eta} = \sigma_{n,s}^{\lambda}\hskip 1pt\varepsilon_{i_1\cdots i_n}^{ \lambda}\, , \label{equ:SigmaS}
\end{align}
where $ \sigma_{n,s}^{\lambda}=0$ for $n<|\lambda|$. The polarization vector $\varepsilon_{i_1\cdots i_n}^{ \lambda}$ is symmetric, transverse, and traceless; for details, see \cite{Lee:2016vti}. 

The helicity-$\lambda$ mode function with $n=|\lambda|$ number of polarization directions satisfies, 
\begin{eqnarray}
{\sigma_{|\lambda|,s}^{\lambda}}&&'' -\frac{2(1-\lambda)}{\eta} \hskip 1pt{\sigma_{|\lambda|,s}^{\lambda}}'\\ +&& \left(k^2+\frac{m^2/H^2-(s+\lambda-2)(s-\lambda+1)}{\eta^2}\right) \sigma_{|\lambda|,s}^{\lambda} =0\, \nonumber .
\label{EoM}
\end{eqnarray}
This admits an exact solution, given by \cite{Lee:2016vti},
\begin{align}
\sigma_{|\lambda|,s}^{\lambda} = {\cal A}_s\hskip 1pt Z_s^{\lambda} (-k\eta)^{3/2-\lambda}H^{(1)}_{i\mu_s}( k|\eta|)\, ,
\label{modefunction}
\end{align}
where $\mu_s$ is defined as  
\beq 
\label{eq:mus}
\mu_s = \sqrt{\frac{m^2}{H^2} - \left(s - \frac{1}{2}\right)^2}.
\eeq 
The normalization coefficients are given by, 
\be
\label{eq:As}
\mathcal{A}_s = e^{i\pi/4}e^{-\pi\mu_s/2}
\ee 
and 
\be 
(Z^\lambda_s)^2 = \frac{1}{k} \left(\frac{k}{H}\right)^{2s-2}(\mathcal{Z}^\lambda_s)^2,
\ee 
with
\begin{eqnarray} 
(\mathcal{Z}^\lambda_s)^2 = && \frac{\pi}{4}\frac{[(2\lambda - 1)!!]^2s!(s-\lambda)!}{(2s-1)!! (s + \lambda)! } \\
&& \cdot\frac{\Gamma(1/2 + \lambda + i\mu_s)\Gamma(1/2 + \lambda - i\mu_s)}{\Gamma(1/2 + s + i\mu_S)\Gamma(1/2 + s - i\mu_s)}. \nonumber 
\end{eqnarray}
The other mode functions can then be obtained iteratively from the recursion relation:
\begin{equation}
\sigma_{n+1,s}^{\lambda} = -\frac{i}{k}\left({\sigma_{n,s}^{\lambda}}'-\frac{2}{\eta} \sigma_{n,s}^{\lambda}\right)-\sum_{m=|\lambda|}^n B_{m,n+1}\hskip 1pt \sigma_{m,s}^{\lambda}\, , \label{moderecur}
\end{equation}
where, 
\begin{equation}
B_{m,n}\equiv \frac{2^n\hskip 1pt n!}{m!(n-m)!(2n-1)!!}\frac{\Gamma[\frac{1}{2}(1+m+n)]}{\Gamma[\frac{1}{2}(1+m-n)]}\, .
\end{equation}

Care should be taken when considering these mode functions, since they are normalized with respect to $\sigma$ with all lower indices. The quantity of physical interest is the two point function of two contracted $\sigma$, i.e.,
\begin{equation}
    \langle \sigma_{i_1 ... i_s} \sigma^{i_s .... i_s}\rangle = \frac{1}{a^{2s}} \langle \sigma_{i_1 ... i_s} \sigma_{i_s .... i_s}\rangle.
    \label{eq:2ptfcnsigmasigma}
\end{equation}
More generally, for $\sigma^{\lambda} _{n, s}$, we are interested in the two-point correlation function,
\begin{equation}
    \langle \sigma^{\lambda} _{n, s} \varepsilon^\lambda _{i_1.... i_n} \sigma^{\lambda} _{n, s}\varepsilon^{\lambda i_1.... i_n} \rangle = \frac{1}{a^{2n}} \langle \sigma^{\lambda} _{n, s} \varepsilon^\lambda _{i_1.... i_n} \sigma^{\lambda} _{n, s}\varepsilon^\lambda_{i_1.... i_n}  \rangle.
\end{equation}
The remaining contraction of polarization vectors can be computed from \cite{Lee:2016vti}
\begin{align}
\varepsilon^{\lambda}_{i_1\cdots i_s}\varepsilon^{\lambda *}_{i_1\cdots i_s} &= \frac{(2s-1)!!(s+\lambda)!}{ [(2\lambda-1)!!]^2s!(s-\lambda)!} ,
\label{eq:contracteps}
\end{align}
where we have used the normalization $\varepsilon^s_{i_1\cdots i_s}\varepsilon^{s *}_{i_1\cdots i_s} = 2^s$ \cite{Lee:2016vti}.

\section{Gravitational Production of Higher Spin Dark Matter}\label{III}

The Higuchi bound $m^2 > s(s-1)H^2$ \cite{Higuchi:1986py,Deser:2001us}  suggests that higher spin fields as realized in nature, insofar as they can be described by a single effective field theory in both the very early universe and in the late universe, should be cosmologically heavy. Guided by past literature \cite{Kolb:1998ki,Chung:2001cb,Chung:1998bt,Chung:1998rq,Kolb:2007vd}, it is logical then to consider gravitational particle production as a genesis mechanism for higher spin dark matter. With this in mind, our first goal is to make a conservative estimate of the gravitational production of higher spin particles in the very early universe. This will serve as a proof of principle of gravitational particle production (GPP) as a genesis mechanism for higher spin dark matter (HSDM).

We will consider only the gravitational production {\it during} inflation, and not the transition between inflation and the radiation dominated phase that is the usual focus of works on GPP of dark matter \cite{Kolb:1998ki,Chung:2001cb,Chung:1998bt,Chung:1998rq, Kolb:2007vd,Kolb:2017jvz,Li:2020xwr}. This simplification is not made for convenience, but rather due to the limited knowledge of higher spin field theories as discussed previously. Our calculation provides a lower bound on the production, suitable for a demonstration that early universe GPP can provide enough higher spin particles to explain the observed DM abundance. We expect a more detailed analysis (e.g., directly from string theory) to change slightly the quantitative relationship between $s$, $m$, and $H$, that leads to the correct relic density, but not the qualitative result.

We focus on HS fields that during inflation are in either the complementary or principal series, defined by Eqs.~\eqref{eq:complseries} and \eqref{eq:princseries} respectively. We make use of the fact that cosmic history is thought to be book-ended by de Sitter phases: cosmic inflation in the first moments and dark energy in the present. The full structure of the higher spin theory is not known in for the intervening time period. However, post-inflation we are left with a collection of non-relativistic particles that simply redshift as matter, and hence we do not need to consider the detailed field theory dynamics. This negates the need to have complete knowledge of higher spin theories.

In general, gravitational particle production occurs when the field mass, including all contributions from quantum and gravitational effects, changes non-adiabatically. The canonical example is the primordial curvature, $\zeta$, which obeys the equation of motion (in exact dS space)
\beq
v_k '' + \left( k^2 - \frac{2}{\eta^2}\right)v_k = 0 .
\eeq
Adiabaticity is violated when $k=\sqrt{2}/|\eta| \simeq a H$, i.e., when a given mode exits the horizon. The resulting particle production can be thought of as Hawking radiation emitted by the de Sitter horizon \cite{PhysRevD.15.2738}, see e.g.~the discussion in \cite{Kinney:2003xf}, and indeed computed using conventional particle production methods \cite{Lapedes:1977ip,Nakayama:1987dh}.  Alternatively, the equation of motion can be solved exactly at all times as a function of $k\eta$, see e.g.~\cite{Baumann:2018muz}.

For a massive scalar field the effective mass is similarly given by \cite{Baumann:2018muz}
\begin{equation}
    \omega_k ^2 = k^2 - \frac{2- (m/H)^2}{\eta^2} .
\end{equation}
For $m/H \gg 1$, adiabaticity is violated at $k\simeq \mu/|\eta|=\mu a H$ where $\mu\equiv m/H$ \cite{Li:2020xwr}. The scale $k_* = \mu a H$ defines an effective horizon, and all modes which have $k>k_*$ at the end of inflation, i.e. which have exited this effective horizon, have undergone particle production during inflation. 

This is the principle behind superheavy dark matter. The total energy density in a massive scalar can be computed as
\begin{equation}
    \rho_\varphi = m^2 _\varphi  \langle \delta \varphi^2 \rangle \simeq m^2 _\varphi \int _{0} ^{k_*} {\rm d}^3 k \, |\delta \varphi_k|^2 ,
    \label{rhoinfl}
\end{equation}
corresponding to a particle number, $n=\rho_\varphi/m_{\varphi}$,
\begin{equation}
    n = m_\varphi \int _{0} ^{k_*} {\rm d}^3 k \, |\delta \varphi_k|^2.
\end{equation}
The phenomenology of superheavy dark matter is thus determined by the dark matter mass and the energy scale of inflation. Due to the decay outside the horizon of the heavy field, along with the phase space suppression in the limit $k\rightarrow 0$, the integral is dominated by the contribution from the upper bound. The number density scales with $H^3$, where $H$ is the Hubble constant during inflation.  The relic density can be tuned to the observed value by tuning the mass $m_{\rm DM}$ as a function of $H$; for the canonical super-heavy dark matter, one finds $m_{\rm DM}\gtrsim H$ leads to the observed density \cite{Kolb:1998ki,Chung:2001cb,Chung:1998bt,Chung:1998rq,Kolb:2007vd}.

The generalization of this to the energy density of a bosonic higher spin field $\sigma$ produced gravitationally during inflation is given by,
\begin{equation}
\label{eq:rhosigma}
    \rho_{\sigma} \simeq m^2 \langle \sigma^2 \rangle = m^2 \sum_{n,\lambda} \varepsilon^{\lambda}_{i_1\cdots i_n}\varepsilon^{\lambda *}_{i_1\cdots i_n}\int {\rm d}^3 k  \frac{|\sigma^\lambda _{n,s}|^2}{a^{2n}},
\end{equation}
where the sum is over all values of $n$ and $\lambda$, and the contraction of polarization vectors is given by Eq.~\eqref{eq:contracteps}.

To evaluate this we begin from the equation of motion Eq.~\eqref{EoM}. To put this in a more familiar form, we rescale $\sigma$ in  Eq.~\eqref{EoM} as $\hat{\sigma}=\eta^{1-\lambda} \sigma$. This removes the first-derivative term, leading to,
\beq 
{{\hat{\sigma}}^\lambda_{|\lambda|,s}} {}'' + \left(k^2 - \frac{s(s-1) - \mu_s^2}{\eta^2}\right){\hat{\sigma}}^\lambda_{|\lambda|,s} =0 ,
\eeq 
which describes a simple harmonic oscillator with a time-dependent mass. We can see that the $\lambda$ dependence has cancelled identically, and the effective mass is given by,
\beq 
\label{eq:meff}
m^2 _{\rm eff}=\mu_*^2 H^2 \equiv (s(s-1) - \mu_s ^2)H^2,
\eeq 
where we define the quantity $\mu_*\equiv s(s-1) - \mu_s ^2$, with $\mu_s$ given by Eq.~\eqref{eq:mus}.

In analogy with a massive scalar, adiabaticity is maximally violated  ($|\dot{\omega}_k|/\omega_k^2$ is peaked) when $k= \sqrt{2} \mu_* a H \simeq \mu_* a H $. Thus, we take the UV cutoff of the energy density integral to be this scale of adiabaticity violation. Incidentally, this cutoff is also the dividing line between relativistic and non-relativistic particles. Thus, independent of discussions of adiabaticity violation, this approach is equivalent to  considering only those perturbations which become non-relativistic already during inflation and thus can be treated as non-relativistic at all times following inflation.

In principle, Eq.~\eqref{eq:rhosigma} involves the sum over all excitations of the spin-$s$ field $\sigma$. However, the dominant excitation of a spin-$s$ field near the Higuchi bound is $\sigma^0 _{s,s}$. This can be seen qualitatively from Eqs.~\eqref{modefunction} and \eqref{moderecur}. Eq.~\eqref{modefunction} will have a maximum value at $\lambda = 0$ and all other modes with $\lambda > 0$ will be suppressed in comparison. It can then be seen from Eq.~\eqref{moderecur} that the dominant contribution will be the $n=s$ state, due to the contribution from $B_{m,n}$. Curiously, this is the same mode which is considered in the `cosmological collider physics' program \cite{Arkani-Hamed:2015bza, Chen:2009we, Chen:2015lza}, leading to the characteristic Legendre polynomial angular dependence of 3-point functions \cite{Arkani-Hamed:2015bza,Lee:2016vti,Alexander:2019vtb}.

Given the subdominance of all other modes, we approximate the full density as that which comes from $\sigma^0_{s,s}$. We use the recursion relation Eq.~\eqref{moderecur} to explicitly numerically calculate the mode functions, $\sigma^0_{n,s}$. We then calculate the density of excitations at the end of inflation using
\beq 
\rho_{\sigma^0_{s,s}} (t_i) = m^2 \frac{(2s-1)!!}{s!}\int_0^{a \mu_* H}{\rm d}^3 k \, a^{-2s} |\sigma^0_{s,s}|^2 ,
\label{eq:0ssrelic}
\eeq 
where $t_i$ denotes the end of inflation, and $\mu_*$ is given by Eq.~\eqref{eq:meff}. The factorial prefactors come from evaluating the contraction of $\lambda=0$, $n=s$, polarization vectors. 

The integral is again dominated by the contribution from the upper bound, and in the numerics that follow we make use of this approximation. Further, since $\mu_* \gtrsim s > 1$, by the Higuchi bound, the DM density is dominated by particles that are still sub-horizon at the end of inflation $(k/a<H)$ and those which re-enter the horizon immediately following inflation $k/a \sim H$. Since $m/H$ is greater than $1$ for all subsequent times, we can approximate these particles as non-relativistic for all times post-inflation. From this we define the present DM density as,
    \begin{equation}
        \rho_{\rm today} = \frac{\rho_{\rm inflation}}{a(t_i)^3},
        \label{eq:densitytoday}
    \end{equation}
where $\rho_{\rm inflation}$ is defined by the integral Eq.~\eqref{eq:0ssrelic}, and $a(t_i)$ is the scale factor of the universe at the end of inflation (we normalize $a=1$  today).

The acceptable parameter space for our HSDM model will be that for which the relic density Eq.~\eqref{eq:densitytoday} matches the observed dark matter abundance. The observed dark matter density is given by $\rho_{\rm DM0}= 3 m_{pl}^2 H_0^2 \Omega_{\rm CDM}$ where $\Omega_{\rm CDM}$ is the abundance observed to be  $\Omega_{\rm CDM}h^2\simeq 0.12$ \cite{Aghanim:2018eyx},  and $H_0 \equiv 100 h \, {\rm km/s/Mpc}=2.13h\times 10^{-33} {\rm eV}$ with $h\simeq0.7$. From this we find $\rho_{\rm DM0}$ evaluates to $\rho_{\rm DM0}=3.95\times10^{-11} {\rm eV}^4$. Meanwhile, the redshift factor in Eq.~\eqref{eq:densitytoday} can be simplified by expand $a(t_i)$ as a ratio of redshifts,
\be 
a(t_i) = \frac{1 + z_{eq}}{1 + z_i}\frac{1}{1 + z_{eq}},  
\ee
where eq refers to matter-radiation equality. We have $z_{eq} \simeq3400$  and $(1 + z) \propto T$ for $z\gtrsim z_{eq}$.  Taking $T_{eq} \simeq0.8 \rm{eV} \simeq1 {eV}$ and instant reheating $T_{re} \simeq (g_* \pi^2/30)^{-1/4} \sqrt{ Hm_{pl}}$, with $g_* \sim 100$,  the above becomes $a(t_i) = 1.43 \times 10^{-18} \sqrt{\frac{\rm eV}{H}}$. 

Putting things together, we find that the DM density after inflation must satisfy,
\be
\rho(t_i) = 1.17 \times 10^{37} \left( \frac{H}{\rm eV}\right)^{3/2} {\rm eV}^4,
\ee
with $\rho(t_i)$ given by Eq.~\eqref{eq:0ssrelic}. 

To gain intuition as to the range of masses that can provide the correct relic density, we note the peculiar case of the complementary series, which occupies a narrow range of masses just above the Higuchi bound. At the saturation limit of the Higuchi bound, we approach partial masslessness and states become gauge redundancies \cite{Baumann:2017jvh}. To avoid this we deform away from the Higuchi bound by a small amount $2\delta$ and consider $m^2/H^2 = s(s-1) + 2\delta$. In this limit, $\mu_s$, Eq.~\eqref{eq:mus}, becomes
\be
\mu_s \rightarrow \frac{i}{2}\left( 1 - 4\delta \right). 
\eeq
One can appreciate from the above that $\mu_s$ is purely {\it imaginary}. This is a feature of the complementary series Eq.~\eqref{eq:complseries}, which corresponds to $0 < \delta < 1/4$. It follows that the exponential suppression, which one might anticipate for excitations of a heavy field, and is encoded in Eq.~\eqref{eq:As}, becomes a phase factor -- i.e., is not a suppression at all. 

\begin{figure}[h!]
    \includegraphics[width=0.49\textwidth]{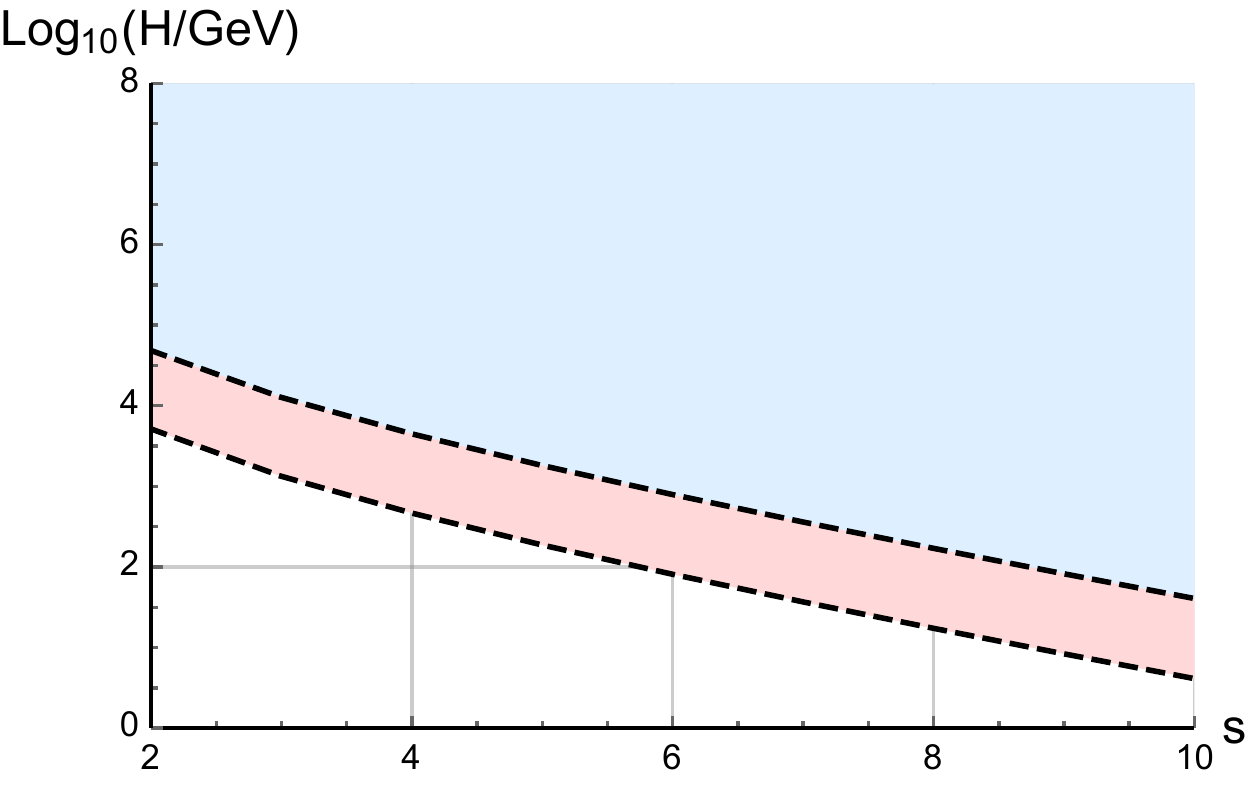}
    \caption{Regions of of $\{s,H \}$ values that give the observed density of higher spin dark matter, with mass in the complementary series (pink) or principal series (blue). The dashed lines denote the boundaries of the complementary series, i.e.,  assuming a mass that differs from the Higuchi bound by a fractional difference $\delta = 0.001$ and $\delta = 0.2499$ as the lower and upper bounds, respectively. The blue region extends up to Planck scale, $H=M_{\rm pl}$.}
    \label{fig:relic}
\end{figure}

In Fig.~\ref{fig:relic} we illustrate the values of $s$ and $H$ for which the correct relic density of particles is produced. By varying the mass, points in the colored regions can achieve the correct relic density. The colors pink and blue denote masses in the complementary series and principle series respectively.

 The lower and upper dashed lines of Fig.~\ref{fig:relic} denote the lower edge of the complementary series (the Higuchi bound) and upper edge of the complementary series. The parameter space below the lower edge is ruled out: the DM particle mass cannot be lower than the Higuchi bound, and decreasing $H$ while leaving $m/H$ fixed will lead to an underproduction of DM during inflation. The upper bound of this band represents the upper limit on particles with masses in the complementary series. If $m/H$ and $s$ remain fixed, increasing $H$ will lead to an overproduction of the DM.  A simple way out is to increase $m/H$, leaving the complementary series, and thereby generating an exponential suppression of the amount of DM, which is denoted by the light blue portion of Figure \ref{fig:relic}.

\begin{figure}[h!]
    \includegraphics[width=0.49\textwidth]{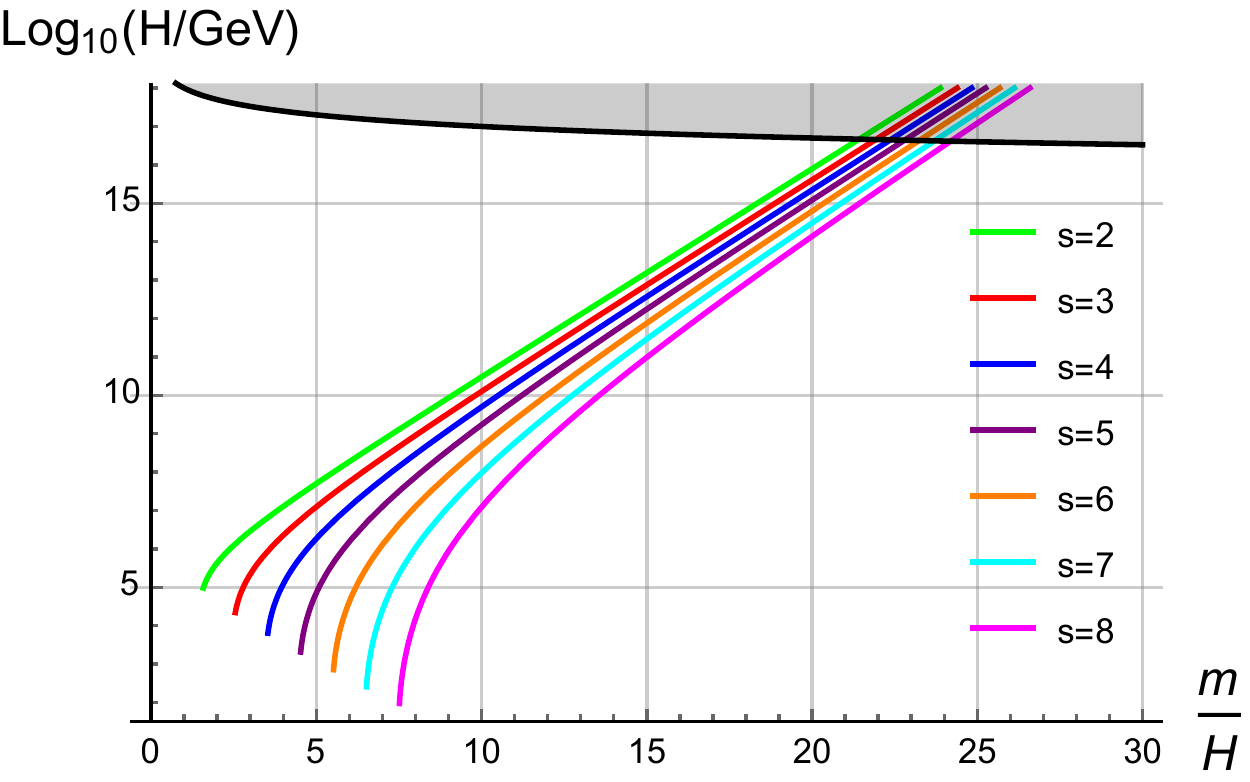}
    \caption{Sample parameter space for particles in the principal series. Lines of fixed $s$ indicate allowable $H$ values as a function of the HSDM mass. Each curve is truncated at the lower boundary of the principal series, i.e., the top dashed line of Fig.~\ref{fig:relic}. The black curve corresponds to masses $m$ equal to the Planck mass, $m=M_{\rm pl}$, and the grey region is super-Planckian masses $m > M_{\rm pl}$.}
    \label{fig:principal}
\end{figure}

By considering masses slightly in the principal series we are able to obtain the correct relic density over the entire blue region of Fig.~\ref{fig:relic}. This imposes a relation between $m$ and $H$, at fixed values of $s$, as shown in Fig.~\ref{fig:principal}. From left to right, fixed curves from $s=2-8$, respectively show that for any spin, there is a continuous range of allowed $H$ values with increasing mass. The allowed parameter space is bounded by the Planck mass, above which is forbidden and denoted by the grey shaded region. Although we have only explicitly shown up to $s=8$, one can appreciate that there is a much larger space where solutions will be present, up to $m/H = 10^{18}$ and correspondingly increasing spin. This allows for a wide range of $H$ and $s$ values, which makes HSDM amenable to a variety of inflation models.

\section{Directional Direct Detection}
\label{IV}

Direct detection is a prominent detection strategy for dark matter. In this approach, one hopes to observe nuclear recoil events generated by scattering of incoming dark matter particles. As realized early on \cite{Spergel:1987kx}, a signature prediction of the motion of the earth through the enveloping dark matter halo is a preferred direction of nuclear recoil events, suggesting an approach known as {\it directional direct detection} \cite{Mayet:2016zxu}. More recently, it has been demonstrated \cite{Catena:2017wzu, Catena:2018uae, Bozorgnia:2011vc} that the angular dependence of the directional direct detection signal can distinguish between spin-$0$, and spin-$1/2$, and spin-$1$ dark matter. It is logical, therefore, to consider the directional direct detection signature of higher spin dark matter.

Any such direct detection signal is premised upon an interaction of dark matter with the standard model. As discussed in Sec.~\ref{II}, while higher spin interactions are naively strongly constrained, this is relaxed for massive higher spins. The interactions of massive higher spin fields can be understood as a low energy effective field theory, with a UV completion given by string theory.  The possible interactions have been studied in detail and enumerated in e.g.~\cite{Buchbinder:2017nuc,Buchbinder:2018wzq,Buchbinder:2018wwg,Gates:2019cnl}.

We will consider a simple interaction between a higher spin boson and a standard model nucleus, eg., Xenon, that we model as a Dirac fermion. We construct an interaction through a derivative coupling of the higher spin boson to the fermion vector current. We consider the low energy effective interaction Lagrangian, 
\beq 
\mathcal{L}_{int} = \frac{g_s}{\Lambda^s} \partial_{\mu_1...\mu_s}(\bar{\psi}\gamma^\mu\psi)\sigma_{\mu}^{\mu_1...\mu_s} + \mathrm{h.c.},
\eeq 
where $\Lambda$ is a UV scale that corresponds to the cutoff of the massive higher spin theory, and $g_s$ is a coupling constant. Note that here $\sigma$ refers to a spin-$(s+1)$ field, rather than spin-$s$. We compute the nuclear recoil scattering cross section in App.~\ref{app:crosssec}. We obtain,
\begin{eqnarray}
\frac{d\sigma}{dE_r} \simeq \frac{m^3}{\pi v^2}\frac{1}{(2s+1)} && \frac{g^4_s}{\Lambda^{4s}}\Big[p^{\prime s}P_s(\hat{k}^\prime \cdot \hat{p}^\prime)+ k^sP_s(\hat{k}^\prime \cdot \hat{k})\Big]^2\nonumber\\ &&\cdot\Big[p^sP_s(\hat{k}\cdot \hat{p})\Big]^2 
\cdot \frac{32k^2}{(p\cdot k)^2},
\label{crosssection}
\end{eqnarray}
where $p$ and $p^\prime$ and $k$ and $k^\prime$ refer to the ingoing and outgoing momenta of the SM nucleus and DM, respectively, $v$ is the relative velocity of the incoming DM particle, and $P_s$ are the Legendre polynomials. The details of the computation of the cross section can be found in App.~\ref{app:crosssec}. 

The quantity relevant to direct directional detection is the double differential rate of nuclear recoil events \cite{Catena:2017wzu, Catena:2018uae}. This is given in standard notation by 
\beq 
\frac{d^2R}{dE_Rd\Omega} = \kappa_{\rm{DM}}\int d^3\mathbf{v}\delta(\mathbf{v}\cdot\mathbf{w} - w_q)f(\mathbf{v} + v_\oplus(t))v^2\frac{d\sigma}{dE_R}. 
\eeq
Here, $v$ is the relative velocity of the incoming DM particle relative to the target nucleus, $\kappa_{\rm{DM}}$ is related to the local halo density of DM near the earth ($\rho_{\rm{DM}} \simeq0.3$ GeV/cm$^3$), $f$ is the velocity distribution of DM in the galactic halo, which is dependent on $\textbf{v}$ as well as $v_\oplus$, the time dependent Earth velocity in the galactic rest frame. We also define \textbf{w} as a unit vector pointing in the direction of nuclear recoil, $w_q = q/(2\mu)$. We will follow the methods in \cite{Catena:2018uae}, outlined below, to obtain an expression for our particular case. 

Following \cite{Catena:2018uae}, we assume a Maxwell-Boltzmann velocity distribution $f(v)$, truncated at the galactic escape velocity, which we take to be $v_{esc} = 544 $ km/s. The most probable speed is $v_0 = 220 $ km/s, which is the circular speed of the local standard of rest. This can be written as 
\beq 
f(\Vec{v}) = \frac{1}{N}e^{-(\Vec{v} + \Vec{v}_e^2)/v_0^2} ,
\eeq 
where 
\beq 
|\Vec{v} + \Vec{v}_e| \leq v_{esc}.
\eeq 
Here $\Vec{v}_e$ is the Earth velocity in the galactic rest frame, which is generally a function of time, but for simplicity can be taken to be its constant magnitude, $\Vec{v}_e = 232$ km/s. The overall normalization $N$ is a constant given by 
\beq 
N = \pi v_0^2\left[\sqrt{\pi}v_0\,\mathrm{erf}\left(\frac{v_{esc}}{v_0}\right) - 2 v_{esc}e^{-v_{esc}^2/v_0^2}\right]. 
\eeq 
We can then break up the velocity integral into two parts: 
\beq
\begin{split}
\int \mathrm{d}^3v  = \int_{v_{min}}^{v_{esc}- v_e} \mathrm{d}v v^2 \int_{-1}^{1}
\mathrm{dcos}\theta \int_0^{2\pi}\mathrm{d}\phi \\
+ \int_{v_{esc} - v_e}^{v_{esc} + v_e} \mathrm{d}v v^2 \int_{-1}^{\frac{v_{esc}^2 - v^2 - v_e^2}{2vv_e}}\mathrm{dcos}\theta \int_0^{2\pi}\mathrm{d}\phi. 
\end{split}
\eeq 
For simplicity here we focus on the first term; the resulting spin and angular dependence is the same in both cases. We make use of the identity,
\beq 
\delta\left(\Vec{v}\cdot \hat{q} - \frac{q}{2\mu}\right) = \frac{\delta(v - \bar{v})}{|\hat{v}\cdot \hat{q}|}, 
\eeq 
where 
\beq
\bar{v} =  \frac{q}{2\mu(\hat{v}\cdot \hat{q})},
\eeq 
where $\hat{v}$ is a unit vector in the direction of the incoming dark matter velocity, $\hat{q}$ is a unit vector in the recoil direction, and $\mu$ is the reduced mass of nucleon-DM system, $\mu = \frac{m_Nm_{\rm DM}}{m_N + m_{\rm DM}}$. Then, performing the integral over the Dirac delta function we obtain,
\begin{eqnarray}
\frac{d^2R}{dE_R d\Omega} = &&\frac{\rho_{\rm{DM}}}{2N\pi m m_N}\int_{-1}^{1}
\mathrm{dcos}\theta \int_0^{2\pi}\mathrm{d}\phi \frac{ \bar{v}^4}{|\hat{v}\cdot \hat{q}|} e^{-(\bar{v} + v_e)^2/v_0^2} \nonumber \\
&&\cdot\frac{d\sigma}{dE_R}(\bar{v})\Theta(\bar{v} - v_{min})\Theta((v_{esc} - v_e) - \bar{v}),
\label{eq:R1}
\end{eqnarray}
where we have taken the dark matter mass to be $m$. Now let us define the angles  $\alpha$, $\beta$, $\theta$, and $\phi$, such that 
\begin{eqnarray}
\hat{q}&& = (\mathrm{sin}\alpha \mathrm{cos}\beta, \mathrm{sin}\alpha \mathrm{sin}\beta, \mathrm{cos}\alpha), \\
\hat{v}&& = (\mathrm{sin}\theta \mathrm{cos}\phi, \mathrm{sin}\theta \mathrm{sin}\phi, \mathrm{cos}\theta)
\end{eqnarray}
 and 
\beq 
\hat{v}^\prime = \frac{v}{v^\prime}\hat{v} + \frac{q}{m v^\prime}\hat{q},
\eeq 
where $\hat{v}$ is the direction of the incoming DM particle, $\hat{q}$ is the direction of nuclear recoil, $\hat{v}^\prime$ is the direction of the outgoing HS DM particle, and $\hat{p}$ is the unit vector in the direction of the incoming nucleus, which we will take to be in the $\hat{z}$ direction. One can deduce $v^\prime$ from conservation of momentum.

Substituting into Eq.~\eqref{eq:R1} the cross section Eq.~\eqref{crosssection}, we find the double differential recoil rate,
\beq 
\begin{split}
\frac{d^2R}{dE_Rd\Omega} \simeq\frac{32\rho_{\rm{DM}} g_s^4 m_N^2}{2N(2s + 1)\pi^2\Lambda^{4s}m}\int d\Omega' f_{\rm SI}(v)\\
\cdot\Big[q^s P_s(\hat{v}^\prime \cdot \hat{q}) + m^s \bar{v}^s P_s(\hat{v}\cdot\hat{v}^\prime) \Big]^2 
\cdot\Big[p^s P_s(\hat{p}\cdot \hat{v})\Big]^2, 
\end{split}
\eeq 
where $d\Omega'$ is the integration over incoming momenta, and $f_{\rm SI}(v)$ is the spin-independent contribution to the scattering rate, given by,
\beq 
\begin{split}
f_{\rm SI}(v) =  \frac{\bar{v}^4}{|\hat{v}\cdot \hat{q}|(p\cdot \bar{v})^2}
\cdot e^{-(\bar{v}^2 + v_e^2+ 2\bar{v}v_e \cos\theta)/v_0^2}\\
\cdot\Theta(\bar{v} - v_{min})\Theta((v_{esc} - v_e) - \bar{v}).
\end{split}
\eeq 
Finally, in the limit that the dark matter mass is larger than the momentum transfer, i.e., $m \gg q$,  the term proportional to $m \bar{v}$ in the second line will be dominant over the first.

The final result is then given by,
\begin{equation}
\frac{d^2R}{dE_Rd\Omega}= \frac{{\cal N}}{\Lambda^{4 s}} \int d\Omega'  f_{\rm SI}(v) \Big[m^s \bar{v}^s P_s(\hat{v}\cdot\hat{v}^\prime) \Big]^2\Big[p^{s} P_s(\hat{p}\cdot \hat{v})\Big]^2,
\label{eq: Recoils}
\end{equation} 
where the prefactor ${\cal N}$ is defined as,
\begin{equation}
    {\cal N} = \frac{32\rho_{\rm{DM}} g_s^4 m_N^2}{2N(2s + 1)\pi^2m} .
\end{equation}
Here, note the distinct dependence on the Legendre polynomials, $P_s$. The angular dependence in $f_{\rm SI}(v)$ is spin-independent and is present in conventional WIMP models. 

The angular dependence on the Legendre polynomials, one the other hand, is a direct consequence of considering higher spins. This generalizes previous works (see e.g. \cite{Catena:2017wzu, Catena:2018uae, Bozorgnia:2011vc, Bozorgnia:2016qkh}) that considered the imprints of spin-0, 1/2, and 1 DM in the double differential recoil rate,leading to ring-like features. Our HSDM model modulates the bosonic signal further, with an added angular dependence on Legendre polynomials, $P_s$, due to scattering of a spin-$s+1$ boson.

This avenue for direct detection is complementary to work that has been done within the `cosmological collider physics' program, which predicts that higher spin particles produced during inflation will leave behind a distinctive signature in the cosmic microwave background, proportional to Legendre polynomials \cite{Chen:2009we, Lee:2016vti, Alexander:2019vtb}. An implication of our results is the potential for a dual `smoking gun' signature and relationship between CMB experiments and dark matter direct detection experiments. We see that the angular dependence on Legendre polynomials in the CMB non-gaussianity is mirrored by the appearance of Legendre polynomials in the double differential recoil rate. 

\section{Other Observable Windows}
\label{V}
Aside from direct detection, one might wonder what signature higher spin dark matter may leave at the other pillars of dark matter detection: collider production, and indirect detection (e.g., at the galactic center \cite{Hooper:2010mq}). To this end, we consider the other possible interactions that could couple higher spin dark matter candidate to the Standard Model.

The simplest possibility is to directly couple a higher spin boson to the Standard Model Higgs boson. For example, an interaction of the form,
\begin{equation}
    {\mathcal L}_{\sigma HH} = \frac{1}{\Lambda^{s-1}}\sigma^{(s)}\partial^{(s)} |H|^2,
\end{equation}
where $\partial^{(s)}$ denotes $s$ number of derivatives. From this interaction, a high energy Higgs boson, e.g. generated at a collider, could radiate a spin-$s$ boson $\sigma$.  At quadratic order in $\sigma$, one could have a $\sigma^2 |H|^2$ interaction, allowing $2\leftrightarrow2$ scattering of the Higgs and HS boson.
 
The field content of the standard model is not restricted to bosons, and nor are higher spin field theories. Higher spin {\it fermions} are interesting in their own right, and may serve as their own dark matter candidate. Unfortunately, a relic density computation as has been done here is not readily repeated for fermions, since the exact solution of HS fermions in dS space is not known. Nonetheless, one may expect a higher spin fermon dark matter candidate to couple to the standard model fermions via 4-fermion interactions, e.g., of the form,
\begin{equation}
    {\mathcal{L}}_{4f} = \frac{1}{\Lambda^2} \bar{\Psi}^{\mu_1 ... \mu_s} \Psi_{\mu_1 ...\mu_s} f \bar{f}   ,
\end{equation}
where $\Psi$ is a spin-$s+1/2$ fermion, and $f$ is a standard model fermion. Such an interaction allows for signatures at precision electroweak experiments, and via annihilation of HS fermions into standard model fermions, a signature in the galactic center \cite{Hooper:2010mq}.

Finally, there is the possibility that the higher spin bosons and higher spin fermions could be organized into multiplets, that is, into {\it super}-multiplets of a supersymmetric theory of higher spin fields \cite{Kuzenko:1993jq,Kuzenko:1993jp,Gates:1996my} (for recent work, see e.g.~\cite{Gates:2010td,Gates:2011qa,Gates:2011qb,Gates:2013ska,Gates:2013tka,Gates:2017hmb,Buchbinder:2017nuc,Buchbinder:2018wwg,Buchbinder:2018gle,Buchbinder:2018wzq,Gates:2019cnl,Buchbinder:2019esz,Buchbinder:2020yip}). Supersymmetry constrains both the spectrum of the theory and the interactions, as discussed in a cosmological context in \cite{Alexander:2019vtb}. Even if supersymmetry is broken at a high scale, one would expect some remnant of this structure to remain at energies accessible by terrestrial experiments. 

The cosmological collider analysis of higher spin supersymmetry \cite{Alexander:2019vtb} revealed correlated signals, with the usual $P_s(\cos \theta)$ angular dependence accompanied by superpartner contributions that scale as $P_{s+1}(\cos \theta)$ and $\sum_m P^m _s(\cos \theta)$. It will be interesting to compute the dual signal in directional direct detection.

\section{Discussion}\label{VI}

In this work we have considered gravitationally produced massive higher spin particles as a model of dark matter. We have shown that there is a wide range of parameter space for superheavy particles with $s>2$ for which the correct relic density of dark matter is produced. We have also explored a potential directional direct detection signature, showing that there is distinctive spin dependent angular dependence in the double differential recoil rate. This enters in the form of Legendre polynomials and is complementary to the `cosmological collider' signature in the cosmic microwave background.

This opens up opportunities to explore a wide range of new models and parameter spaces to aid in the search for dark matter. We have mentioned several possibilities, but there is certainly much that is still unknown. In future work we will perform a more rigorous numerical exploration of our results in the context of directional direct detection, in order to show more explicitly the impact of the higher spin angular dependence in the double differential recoil rate. It would also be of interest to build a similar model for fermionic HSDM and build connections to HS supersymmetry. We leave these explorations and others to future work.

{\bf Comment added:} During the final preparation of this manuscript we became aware of recent work \cite{Criado:2020jkp} with thematic overlap to this paper. The effective field theory approach taken there is promising and may yield further directional direct detection signatures in addition to that considered here. There is no overlap in the content of the papers.

\acknowledgements 
We thank Sylvester James Gates, Jr., Rocky Kolb, Konstantinos Koutrolikos, and Andrew Long, for useful discussions.

\appendix
\section{Matrix Element Calculation}
\label{app:crosssec}
We here show technical details of the differential cross section used in Section \ref{IV}. To calculate the differential cross section for a HSDM particle scattering off a SM fermion, there are several subtleties that must be considered. The vertex factor for scattering with $\sigma^{0} _{s,s}$ is given by 
\beq 
V= -i \frac{g_s}{\Lambda^s}\left[\sum_{n=0}^{s} {s\choose n}k_1^{i_1...i_{s-n}}k_2^{i_n...i_s}\right]\gamma^\mu,
\eeq 
where we have considered all possible combinations of derivatives on $\bar{\psi}\psi$, and $k_1$ and $k_2$ correspond to the momenta of the standard model nucleons. We assume that the incoming nucleon is at rest, therefore the only terms that contributes to the vertex will be those which have either $k_1^s$ or $k_2^s$. The helicity state of the higher spin particle is  $\lambda$. Thus, this expression can be simplified as 
\beq 
V= -i \frac{g_s}{\Lambda^s} (k_{1_{i_1...i_s}} + k_{2_{i_1...i_s}})\gamma^\mu.
\eeq 
Each higher spin external leg carries a factor of the spin-$s+1$ polarization tensor $\epsilon_\mu^{[\lambda'']i_1...i_s}(k_3)$. This can be decomposed into a spin-$1$ component and a spin-$s$ component as follows:
\beq 
\epsilon_\mu^{[\lambda'']i_1...i_s}(k_3) = \epsilon_\mu^{[\lambda^\prime]}(k_3)\epsilon^{[\lambda]i_1...i_s}(k_3),
\eeq 
where $\lambda^\prime = -1,0,1$ and $\lambda=-s...s$ are the possible helicity states of the spin-1 and spin-$s$ components, respectively. Lastly, note that generally, working in an expanding background will lead to additional factors of the scale factor, $a(t)$. However, for the remainder of the calculation we normalize $a(t) = 1$, to account for the insensitivity of particle physics experiments today to the previous expansion of the universe. The matrix element can easily be found in analogy with standard QED computations, and with the use of the relation
\beq 
\hat{q}_{i_1}...\hat{q}_{i_s}\epsilon^\lambda_{i_1...i_s} \equiv \mathcal{E}^\lambda_\lambda(\theta, \phi)P^\lambda_s(\cos\theta), 
\eeq 
where $\cos\theta = \hat{\textbf{q}}\cdot \hat{\textbf{k}}$, $\cos\phi = \hat{\textbf{q}}\cdot\epsilon$ and $\mathcal{E}^\lambda_\lambda$ and $P_s^\lambda$ are the transverse and longitudinal parts of the spherical harmonics, respectively \cite{Lee:2016vti, Baumann:2017jvh}, where for the $\lambda = 0$ modes we simply have  $\hat{q}_{i_1}...\hat{q}_{i_s}\epsilon^\lambda_{i_1...i_s} = P_s(\cos\theta)$. Then, we find the matrix element to be 
\beq 
\begin{split}
|\mathcal{M}|^2 =\frac{g^4_s}{\Lambda^{4s}}  \Big[p^{\prime s}P_s(\hat{k}^\prime \cdot \hat{p}^\prime)+(p+k)^s P_s(\hat{k}^\prime \cdot \hat{(p+k)})\Big]^2\\
\cdot\Big[p^sP_s(\hat{k}\cdot \hat{p}) + (p+k)^sP_s(\hat{k} \cdot \hat{(p+k)})\Big]^2\\
\cdot\frac{(2m)^2}{(2p\cdot k)^2} (16 p^2 + 64 p\cdot k + 32 k^2),
\end{split}
\eeq
where $k$ and $k^\prime$ refer to the incoming and outgoing HSDM particle, respectively,$p$ and $p^\prime$ refer to the standard model nucleus, and we note $\hat{(p+k)} \equiv (\vec{p}+\vec{k})/|\vec{p}+\vec{k}|$. From the matrix element, one can find the differential scattering cross section given by 
\beq 
\frac{d\sigma}{dE_R} = \frac{2m}{\pi v^2}\frac{1}{(2J + 1)(2s_\chi + 1)}|\mathcal{M}|^2.
\eeq 
 We will simplify this by noting that in our construction, we assume that the standard model nucleus is stationary and the incoming HS particle has a much larger momentum, $k \gg p$. In this limit, we can also take $k^\prime \cdot (p+k )= k^\prime \cdot k$. Thus, keeping only the relevant dominant terms in k, we obtain for the cross section 
\begin{eqnarray}
\frac{d\sigma}{dE_R} \simeq \frac{m^3}{\pi v^2}\frac{1}{(2s+1)} && \frac{g^4_s}{\Lambda^{4s}}\Big[p^{\prime s}P_s(\hat{k}^\prime \cdot \hat{p}^\prime)+ k^sP_s(\hat{k}^\prime \cdot \hat{k})\Big]^2 \nonumber \\
\cdot && \Big[p^sP_s(\hat{k}\cdot \hat{p})\Big]^2 
\cdot \frac{32k^2}{(p\cdot k)^2}. 
\end{eqnarray}

\bibliographystyle{JHEP}
\bibliography{HSDM-refs}

\end{document}